\documentstyle[aps,epsfig,multicol]{revtex}
\begin{document}
\draft
\title{Interaction of quantum Hall systems with waveguide elastic modes}
\author{D.V.Fil}
\address{Institute for Single Crystals National Academy of Sciences of
Ukraine,
Lenin av. 60 Kharkov 310001 Ukraine\\ e-mail: fil@isc.kharkov.ua}
\date{December 9,1998}
\maketitle
\begin{abstract}
An interaction of non-uniform plane elastic modes of the
waveguide type with monolayer and double-layer quantum
Hall systems is considered.
It is shown, that unlike the case of the
surface acoustic wave propagation, the restriction on
maximal values of the wave vectors for which the
velocity shift can be observed experimentally
does not take place for the waveguide modes.
In case of study of incompressible fractional quantum Hall states,
the effect can be used for measuring
a dependence of the effective magnetic length
on the filling factor
and for observing phase transitions in double-layer systems
under the interlayer distance variation
\end{abstract}
\pacs {PACS numbers: 71.10 Pm, 73.40.Hm, 74.20.Dx}
\begin{multicols}{2}
\section{Introduction}
The interaction of surface acoustic waves (SAW) with quantum
Hall systems was intensively studied both experimentally \cite{1,2}
and theoretically \cite{3,4,5,6,7,8,9} during the last few years.
Measurements of both velocity shift and absorption of SAW's provide
an indirect method of studying the dynamical conductivity of a
two dimensional electron(2DE) gas.
The SAW experiments (together with a study of the
magnetic focussing effect \cite{10,11} and of a temperature
dependence of the d.c. conductivity of the quantum Hall
system \cite{12,13,14}) are considered as an experimental proof for
the composite fermion model \cite{15,16,2}
of the fractional quantum Hall(FQH) effect.

In experiments, the SAW is localized on the surface of the sample, while
the two dimensional electron layer is placed at some depth $d$
(typical values of $d$ are $10^3\div 5\cdot 10^3 \AA$).
Therefore the study of a momentum ($q$) dependence of the conductivity is
restricted  by the inequality  $q< q_0$, where $q_0$ is of order of $d^{-1}$.

The measurements beyond that region are especially important for an
investigation of dynamical properties of incompressible FQH states.
It was shown in Ref.~\cite{17}, that
at filling factors ($\nu $) corresponding to such states
an interaction of acoustic phonon with 2DEs results
in an essential phase velocity shift ($\Delta v$) for phonons at  finite
wave vectors. According to Ref.~\cite{17}, the
function $\Delta v(q)$ oscillates at finite $q$. The period of
oscillation is $\sim \lambda^{-1}_{eff}$, where  $\lambda_{eff}$
is the effective magnetic length. Therefore, the measurements of
$\Delta v(q)$ can be used as a way of an observing
the $\lambda_{eff}(\nu )$ dependence.
It was shown Ref.~\cite{18}, that the measurements of the
velocity shifts of high frequency phonons interacting with a double
layer FQH system can be also used
as a method of a registration of phase transitions between FQH states
described by different quantum numbers. (They are the
generalized Laughlin states, which correspond to the Halperin wave
function \cite{19}. The composite fermion theory for such states was
developed by Lopez and Fradkin \cite{20}.)

The values of the velocity shifts obtained in Refs.~\cite{17,18}
correspond to bulk acoustic waves interacting with
a superlattice of 2DE layers.
For a single monolayer (or
a double layer) system,
the elastic wave should be localized near
the electron layer to provide the values of the effect, which can be
observed experimentally. At large $q$ it is not the case of SAW, which
is localized near the surface. The experimental geometry
required can be realized in layered elastic media, where waveguide
elastic modes may exist. We consider the case of an infinite elastic
medium which incorporates an elastic layer with the acoustic properties
different from the ones for the bulk.
The interaction of non-uniform transverse elastic waves propagating
in such a medium with a quantum Hall system is studied.
We consider the case, for which
the waves vectors and the polarization vectors of the non-uniform modes
are parallel to the
interface boundaries. The velocity of
the transverse sound for the layer is assumed to be smaller then
for the bulk matrix (it is the condition of the existence of such
modes). The piezoelectric interaction mechanism is implied.

Heterostructures $Al_{x}Ga_{1-x}As -GaAs$ used in experimental
studies of double layer quantum Hall systems obey the
elastic properties required.
For instance, the structures $A-B-A-B-A$
and $A-B-A$ (where $A$ - $Al_{x}Ga_{1-x}As$, $B$ - $GaAs$)
were used in Refs.~\cite{21,22}.
Such structures can be approximately described by a model
of a bulk sample of $AlGaAs$ incorporating a layer of $GaAs$.
(One can expect that  the effect
of the surface, the substrate and the thin central layer
(for the $A-B-A-B-A$ structure) can be neglected for the wave vectors
considered). The transverse sound velocity for $GaAs$  is smaller
then for $Al_{x}Ga_{1-x}As$. Therefore, these exist non-uniform
elastic modes localized near the $GaAs$ layer.
As it follows from our results, the values of the velocity shifts
of those modes are within the experimental resolution, while they are
smaller than in the SAW experiments.  We also consider the geometry for
which the heterostructure is places in the matrix of another material.
For the last case the values of the effect are of the same order or higher
than for the SAW.

\section{Matrix elements of interaction}

Let us start from a simple problem of renormalization of the phase velocity
of acoustic phonons interacting with 2DEs. The phonon part of Hamiltonian
has the form
\begin{eqnarray}
H_{ph}=\sum_{\bf q} \omega _q (b^+_{\bf q} b_{\bf q} +{1\over 2}),
\label{1}
\end{eqnarray}
where $\omega _q = v q$ is the phonon spectrum,
$b^+$($b$) are the phonon creation (annihilation)
operators. The Hamiltonian of the electron-phonon interaction is chosen
in the form
\begin{eqnarray}
H_{int} = {1\over  \sqrt{S}} \sum_{{\bf q},k} \int d^2 r
\ g_{q k}\Psi^+_{{\bf r} k} \Psi_{{\bf r} k}
e^{i {\bf q r}}(b_{\bf q} + b^+_{-{\bf q}}),
\label{2}
\end{eqnarray}
where $\Psi^+(\Psi)$
are the electron creation (annihilation)
operators,
$k$ is the electron layer index,
$S$ is the area of the layer,
$g$ are the matrix elements of the interaction.

The velocity shift under such a choice of
$H_{ph}$ ¨ $H_{int}$ is described by the expression
\begin{eqnarray}
{\Delta v\over v}={1\over v q} \sum_{k,k'} g_{q k}^* g_{q k'}
D_{k k'}(q,v q),
\label{3}
\end{eqnarray}
where $D(q,\omega)$ is the electron density-density response
function.
In the composite fermion approach \cite{3,16} an additional statistical
interaction between electrons is introduced, which results in a
modification of the functions $D$ in Eq.~(\ref{3}).

We will use the approach similar to  that of Ref.~\cite{4} for
the calculations of the matrix elements
$g_{q k}$.
The piezoelectric interaction of the elastic wave with the electrons
corresponds to the Hamiltonian
\begin{eqnarray}
H = \sum_{k} \int d^2 r \ e \varphi_{{\bf r} k}
\Psi^+_{{\bf r} k} \Psi_{{\bf r} k},
\label{4}
\end{eqnarray}
where $\varphi$ is the electric potential, generated by the
elastic mode. According to the linear response theory
$\varphi$ in Eq.~(\ref{4}) is
the external potential. Therefore, to calculate its value
one should neglect the influence of the electron system.

We will use the smallness of the piezoelectric interaction
(the explicit expression for the small parameter is given below).
Let us discuss the solution of the elastic problem with the
piezoelectric constant set to zero. Let the interfaces be in the
$x$-$y$ planes with $z=\pm a$ coordinates.
We denote $c_l$,
the transverse sound velocity in the
layer ($|z|<a$), and
$c_b$, the transverse sound velocity in the
bulk matrix ($|z|>a$). The condition $c_b>c_l$ is implied.
For definiteness, let us assume that the crystal symmetry of the bulk
matrix and of the layer is the cubic one, and the elastic moduli $c_{44}$
are the same for the layer and for the matrix,
while they have different densities.
(We use such a special choice of parameters to simplify the
expressions obtained. This choice corresponds approximately
to the parameters for $AlGaAs$ and $GaAs$.)
We consider the case when the axis $x$ is chosen along the
[100] direction
and the $z$ axis - along the [001] direction

Let the wave vector and the displacement vector of the elastic wave are
directed along the $x$ and $y$ axes correspondingly.
The wave equation for the Fourier component of the displacement
$u_y(q,z,\omega)=u$ reads as
\begin{eqnarray}
{\partial^2 u \over \partial z^2 } - (c_{l(b)}^2 q^2 -\omega^2) u =0.
\label{5}
\end{eqnarray}
One can find from the boundary conditions that the function
$u(q,z)$ and $\partial u(q,z)/\partial z$
are continuous ones at the interfaces. The localized solutions of
Eq. (\ref{5}) have the form
\begin{eqnarray}
u_e(q,z)=A_q^e \cases { \cos (\eta q a)
e^{-\lambda q (|z|-a)} &$|z|\geq a$\cr
\cos (\eta q z) &$|z|\leq a$}
\label{6}
\end{eqnarray}
(the even mode),
\begin{eqnarray}
u_o(q,z)=A_q^o \cases {
\sin (\eta q a)
e^{-\lambda q (z-a)} &$z\geq a$\cr
\sin (\eta q z) &$|z|\leq a$\cr
-\sin (\eta q a)
e^{\lambda q (z+a)} &$z\leq -a$}
\label{7}
\end{eqnarray}
(the odd mode).
In Eqs. (\ref{6},\ref{7})
$\lambda =\sqrt{1-v^2/c_b^2}$,
$\eta  =\sqrt{v^2/c_l^2-1}$.
The velocity $v$
for the even mode is given by the
solution of the equation
\begin{eqnarray}
\tan (\eta q a)=\lambda /\eta,
\label{8}
\end{eqnarray}
and for the odd mode  - by the
solution of the equation
\begin{eqnarray}
\cot (\eta q a)=-\lambda /\eta.
\label{9}
\end{eqnarray}
At $q<\pi/(2 a \sqrt{c_b^2/c_l^2-1})$
there exists only one (even) mode.
At larger $q$ other solutions of Eqs.(\ref{6},\ref{7}) emerge
(waveguide modes).

The transverse component of the displacement vector can be rewritten
in terms of $b$ operators:
\begin{eqnarray}
u_{e(o)}(r,z)=\sum_q u_{e(o)}(q,z) e^{i q r} (b_q+b_{-q}^+).
\label{10}
\end{eqnarray}
Inserting $u_{e(o)}(r,z)$ into the expression for the energy of
elastic vibrations
\begin{eqnarray}
E_u={1\over 2}\int d^2 r d z [ \rho(z)
 \left( {\partial u/ \partial t}\right)^2 \cr +
c_{44} \left(\left( {\partial u/ \partial x}\right)^2+
\left( {\partial u/ \partial z}\right)^2\right)]
\label{11}
\end{eqnarray}
and comparing  the result with Eq.~(\ref{1}),
we determine the values of the normalization factors in
Eqs.~(\ref{6},\ref{7}):
\begin{eqnarray}
A_q^e=A_q^o= {1\over \alpha } \sqrt{v\over 2 S c_{44}},
\label{12}
\end{eqnarray}
where $\alpha=(q a (1 + \eta^2) + 1/\lambda)^{1/2}$.

The electric potential can be found from the solution of the
Poisson's equation
\begin{eqnarray}
\Delta \varphi = -{4\pi\over \epsilon}
\beta_{i,j m} \partial_i u_{j m},
\label{13}
\end{eqnarray}
where $u_{j m}$ is the strain tensor,
$\epsilon $ is the dielectric constant, $\beta_{i,j m}$ is
the piezoelectric tensor.
In cubic crystals all nonzero component of
 $\beta_{i,j m}$
(with $i\ne j\ne m$) are equal to $\beta$.

The solution of Eq.~(\ref{13})
under substitution of Eqs.(\ref{10},\ref{12})
can be written in the form
\begin{eqnarray}
\varphi_{r z} = i\chi \sqrt{2\pi v\over S  \epsilon }
\sum_q \gamma_q(z) e^{i q r}
(b_q+b_{-q}^+),
\label{14}
\end{eqnarray}
where $\chi=\beta \sqrt{16\pi /(\epsilon c_{44})}$
is the dimensionless parameter of the piezoelectric interaction
( $ \chi$ is the small parameter of our consideration),
$\gamma_q$ is the structural factor,
which is determined by the boundary conditions on
$\varphi$ and by the type of the elastic mode.
By using  Eqs.~(\ref{2},\ref{4},\ref{14}), we arrive at the
expression
\begin{eqnarray}
g_{q k}=  i \sqrt{2\pi \over
\epsilon } e v^{1/2} \chi \gamma_q(z_k),
\label{15}
\end{eqnarray}
where $z_k$ is the $z$ coordinate of the electron layer.
The explicit expressions for $\gamma_q(z_k)$  will be presented in
Sec.IV, when we specify the geometry of the model.

\section{Response functions in the composite fermion model}

Let us discuss the calculation of the response functions
$D$ for the composite fermion model. At the beginning, consider
the monolayer electron system. The random phase approximation
gives the answer
\begin{eqnarray}
D=(1-D_0 V)^{-1}D_0,
\label{16}
\end{eqnarray}
where $V=2 \pi e^2/\epsilon q$ is
the Fourier component of the Coulomb interaction, and
$D_0$ can be expressed though the polarization operator
($\Pi_{\mu \nu }$) of the electromagnetic field
\begin{eqnarray}
D_0=-{1\over e^2}\Pi_{00}.
\label{17}
\end{eqnarray}

We use the approach which is the Lagrange
formulation of the modified random phase approximation (Ref.~\cite{23})
for the calculation of $\Pi_{00}$.
In Ref.~\cite{23} the Landau Fermi liquid
interaction was taken into account. The $F_1$ constant (see, for example,
Ref~\cite{24}) of that interaction is determined by the ratio
$m^*/m_b$, where $m^*$ is the composite fermion effective mass,
$m_b$ - is the band mass of electrons. It was shown in Ref.~\cite{23},
that taking into account that interaction one obtains the response function,
which satisfy the $f$-sum rule.

The Lagrangian of composite fermions can be written in the form
\begin{eqnarray}
L=\Psi^*(i\partial_t + \mu_F  - a_0 - e A_0 - \cr
{1\over 2 m^*} (i\nabla_i + a_i + e A_i + b_i)^2)\Psi + \cr
{1\over 4\pi \psi} a_0 \epsilon_{i j} \partial_i a_j +
{1\over 2} \xi   b_i^2 ,
\label{18}
\end{eqnarray}
where $a_\mu$ is the potential of the Chern-Simons gauge field,
$A_\mu$ is the electromagnetic field potential,
$\epsilon_{ij}$ is fully asymmetric unit tensor,
$\mu_F$ is the chemical potential,
$\psi$ is the parameter of the model,
which corresponds to the number of the gauge field flux quanta
carrying by the composite quasiparticle
($\psi$ is the even number), $b_i$ is an auxiliary field,
which is introduced for modelling the Fermi liquid
interaction. In Eq.~(\ref{18}) the transverse gauge for the $a$ field
is implied.

The value of $\xi$ can be determined from the condition of
that the Lagrangian (\ref{18}) provides the
Galilean invariant value for the electric current (see,
for example, Ref.~\cite{24}):
\begin{eqnarray}
<{\delta  L\over \delta  A_i}>=
<j_i> - {e n_0\over m^*}<b_i>={m^*\over m_b} <j_i>,
\label{19}
\end{eqnarray}
where $j_i=-(e/m^*)\Psi^*(i\nabla_i +a_i+e A_i)\Psi$,
$n_0$ is the uniform electron concentration.
The mean value of the field $b$
is determined by the equation
\begin{eqnarray}
<{\delta  L\over \delta  b_i}>=0= {1\over e} <j_i> -{n_0\over m^*}<b_i>+
\xi <b_i> .
\label{20}
\end{eqnarray}
The condition of consistency of Eqs.~(\ref{19},\ref{20}) yields
$\xi =n_0/(m^* - m_b)$.

The self-consistent effective field acting on
the composite fermions is
$B_{eff}=B(1-\nu \psi)$ (where $B$ is the external magnetic field).
At filling factors $\nu_f = N/(N\psi\pm1)$
($N$ is an integer value)  $B^f_{eff}=\pm n_0 \varphi_0/N$,
where $\varphi_0$ is the magnetic field flux quantum.
The field $B^f_{eff}$ corresponds to the integer value
$N$  of filled Landau levels. Therefore,
$\nu_f$'s correspond to the incompressible states.
The field $B^f_{eff}$ defines
the effective cyclotron energy
$\omega_{ce}=2\pi n_0/m^* N$
and the effective magnetic length
$\lambda_{eff}=(N/2\pi n_0)^{1/2}$.

To determine the value of $\Pi_{00}$ one can calculate the
quadratic part of the effective action for the electromagnetic
field. To do that we calculate the functional integral over
the fields $\Psi$ in the expression for the partition function of
the system, described by the Lagrangian (\ref{18}). Then we
calculate the functional integral over the fields $a$ and $b$
in the vicinity of the saddle point. Here we present the
final result at $T=0$
\begin{eqnarray}
\Pi_{00}= -{e^2 q^2\over 2 \pi \omega_{ce}} {S_0\over \Delta_1 },
\label{21}
\end{eqnarray}
where
\begin{eqnarray}
&&S_0=\Sigma_0 - p (\Sigma_0(\Sigma_2+N)-\Sigma_1^2),\\
&&\Delta_1=(1-\psi \Sigma_1)^2 -\psi^2 \Sigma_0 (\Sigma _2+N)- p F,\\
&&F=\Sigma_2+N + (\omega/\omega_{ce})^2 S_0
\label{22-24}
\end{eqnarray}
with $p=(m^*-m_b)/(m^* N)$,
\begin{eqnarray}
\Sigma _j=({\rm sgn}(B_{eff}))^j
e^{-x}\sum_{n=0}^{N-1} \sum_{m=N}^\infty {n!\over m!}\cr\times
{x^{m-n-1}(m-n) \over
(\omega / \omega _{ce})^2 - (m-n)^2}
[L_n^{m-n}(x)]^{2-j} \cr \times
[ (m-n-x)L_n^{m-n}(x) + 2x{d L_n^{m-n}(x)\over d x}] ^j.
\label{25}
\end{eqnarray}
In Eq.~(\ref{25}) $x=(q\lambda _{eff} )^2/2$,
$L_n^{m-n}(x)$ is the generalized Laguerre polynomial.

Substituting Eqs. (\ref{15}-\ref{17},\ref{21}) into Eq.~(\ref{3})
we obtain
\begin{eqnarray}
{\Delta v\over v} =  \chi^2 \gamma_{q}^2(z_1) {f_q S_0\over
\Delta_1 - f_q S_0 },
\label{26}
\end{eqnarray}
where $f_q = e^2 q/(\epsilon \omega_{ce})$

The case of the double layer Hall system is considered by a similar way.
Two types of the Chern-Simons fields and the fermion fields are introduced
(they correspond to the two electron layers).
The free part of the Chern-Simons Lagrangian includes non-diagonal
terms with respect to the layer index.
\begin{eqnarray}
L_{CS} = {1\over 4\pi } \epsilon_{ij} a_{0k} M_{k k'}
\partial_i a_{j k'}
\label{27}
\end{eqnarray}
with
\begin{eqnarray}
M={1\over \psi^2 - s^2} \left( \matrix{\psi&-s\cr-s&\psi}\right),
\label{28}
\end{eqnarray}
where $s$ (an integer number) is the number of
the Chern-Simons flux quanta of the type "1", carrying
by the quasiparticle of the type "2" and vice versa
(there exists the interlayer statistical interaction in the
system). Here we specify the case of two equivalent electron
layers.
At $\psi=s$ it is better to rewrite the Lagrangian in terms
of the in-phase and out-of-phase combinations of the fields $a_k$.

The incompressible states for the double layer system correspond to
the filling factors
\begin{eqnarray}
\nu_f = {N\over N(\psi+s)\pm 1}
\label{29}
\end{eqnarray}
(calculated for one layer).
The expressions for $B_{eff}^f$, $\omega_{ce}$,
$\lambda_{eff}$ remain unchanged.
Eq.~(\ref{16}) obeys the matrix form with
\begin{eqnarray}
V={2 \pi e^2\over \epsilon q} \left( \matrix{1&e^{-q d_0}\cr
e^{-q d_0}&1}\right),
\label{30}
\end{eqnarray}
where $d_0$ is the distance between the electron layers.

The expression for the velocity shift is modified to the following
one
\begin{eqnarray}
{\Delta v\over v}= {1\over 2}\chi^2 S_0 f_q [
{(\gamma_{q}(z_1)+\gamma_q(z_2))^2
\over \Delta_+ - f_q E_+ S_0} \cr+
{(\gamma_{q}(z_1)-\gamma_{q}(z_2))^2
\over \Delta_- - f_q E_- S_0}],
\label{31}
\end{eqnarray}
where   $E_\pm =1\pm \text{exp}(-q d_0)$,
\begin{eqnarray}
\Delta_\pm=(1-(\psi\pm s) \Sigma_1)^2 \cr -(\psi\pm s)^2
\Sigma_0 (\Sigma_2+N)
-p F.
\label{32}
\end{eqnarray}

The ground state of the double layer system is described
by the quantum numbers
$\psi$ and $s$.
One can see from Eq. (\ref{29}), that the same
$\nu$  may correspond to different sets of
$\psi$ and $s$ (different phases).
If the correlation between the layers is a small one
(at large $d_0$), the phase with $s=0$ is realized.
Reduction of the interlayer distance may result in a phase transition
into the state with $s\ne 0$ (see, for instance, Ref. \cite{25}).
The transition may be accompanied by the shift of the velocity of
the acoustic wave, interacting with the electrons, as described
by Eq.~(\ref{31}).

\section{Velocity shift for the waveguide elastic modes}

Eqs.  (\ref{26},\ref{31}) define the value of
$\Delta v$  relative the velocity for the systems
without the electron-phonon
interaction. In experiments the value of $\Delta v$ is measured
relative to the velocity at $B=0$. To compare the theoretical
value with the experimental data one should know the value
of $\sigma_{xx}(q,\omega)$
(longitudinal component of the conductivity tensor) at zero
magnetic field. One can use the approximation $\sigma_{xx}\to\infty$ at
$B=0$, but in that case the momentum dependence
$\sigma_{xx}(q,\omega)$ at $B=0$ is neglected.
We propose to measure the difference between the sound velocities
at $\nu =\nu_f$ and $\nu =1$. The case of $\nu =1$ is also described
by  Eqs. (\ref{26}), (\ref{31}), if one sets
$\psi=s=0$, $N=1$ ¨ $m^*=m_b$.

Thus, the value we calculate is given by the expression
\begin{eqnarray}
{\Delta v\over v} = {\Delta v\over v}\Bigg|_{\nu =\nu_f} -
{\Delta v\over v}\Bigg|_{\nu =1}.
\label{33}
\end{eqnarray}

Let us consider the structure  $Al_{0.3}Ga_{0.7}As - GaAs -
Al_{0.3}Ga_{0.7}As$, which incorporates two electron layers
at $z_k=\pm a$ ($d_0=2 a$).
The parameters $\beta$ and $\epsilon$ are implied to be uniform
for the whole system. Then, the boundary conditions for
$\varphi$ require the continuity of the values of
$\varphi(q,z)$ and $\partial \varphi(q,z)/\partial z$ at
the interfaces.

The solution of Eq.~(\ref{13}) for the even mode (Eq.~(\ref{6}))
gives an odd function of $\varphi(z)$, while for the
odd mode (Eq.~(\ref{7})) - an even function of $\varphi(z)$.

At $\nu =1/5$ one can expect the phase transition between
($\psi=4$, $s=0$) and ($\psi=2$, $s=2$) states. We see from
Eq. (\ref{31}), that the transition is accompanied by the
shift of the velocity of the even elastic mode, while the
odd mode does not show such a behavior.

For the even mode the values $\gamma_{q}(z_k)$ are found to be
\begin{eqnarray}
\gamma_{q}(a)= -\gamma_{q}(-a)=
\alpha^{-1} {\cos(\eta q a)\over 1+\eta^2 } \cr \times
\left(
{E_- (\eta^2+\lambda ^2)\over 2(1+\lambda) } - \lambda \right)
\label{34}
\end{eqnarray}

The dependencies  $\Delta v/v$ on the wave vector are
shown in Fig.1 at
$\psi=4$, $s=0$ and $\psi=2$, $s=2$.
The parameters
$\beta =4.5\cdot 10^4$
dyn$^{1/2}$/cm, $\epsilon$=12.5, $n_0=10^{11}$ cm$^{-2}$,
$m_b=0.07 m_e$ ($m_e$ is the bare electron mass), $m^*=4 m_b$,
$c_{44}=6\cdot10^{11}$ dyn/cm$^2$, $\rho_l$=5.3 g/cm$^3$,
$c_b/c_l=1.05$, $d_0=300 \AA$ are used for the calculations.
We see, that at finite $q$ the value of the
velocity shift is within the experimental resolution, while it is
rather small due to the smallness of $\Delta c=(c_b-c_l)/c_l$.

\begin{figure}
\narrowtext
\centerline{\epsfig{figure=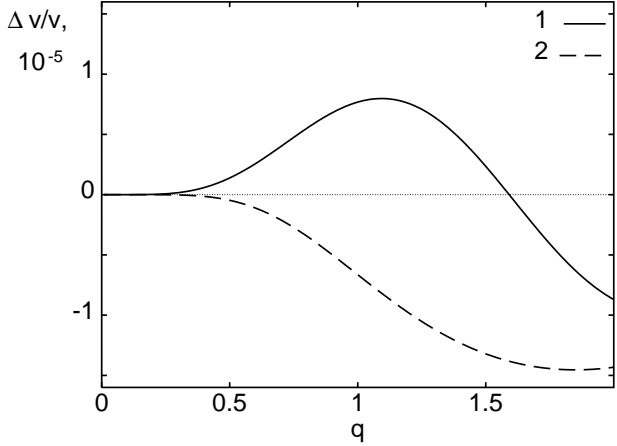,width=8cm}}
\vspace{0.5cm}
\caption{Velocity shifts of the lowest even mode vs the wave vector
for the double layer electron
system at $c_b/c_l=1.05$. 1 - $\varphi =4$, $s=0$; 2 - $\varphi=2$, $s=2$.
$q$ is in $\sqrt{4 \pi n_0}$ units.  }
\label{fig1}
\end{figure}

Now we analyze the geometry for which one could expect larger values
of the effect. Let the heterostructure is placed into the matrix
of another material for which the value of $\Delta c$ is much larger then
for the previous case. For simplicity,
we consider the heterostructure as a uniform acoustic medium
and assume that there are thin screening layers on the interfaces
(then the dielectric and piezoelectric properties of the bulk matrix
are unimportant). The boundary conditions are
$\varphi(z=\pm a)=0$.

If the structure contains  a single electron layer at
$z=0$, then the velocity shift can emerge only for the odd mode.
The effect has a threshold character (at small $q$ the odd mode
does not exist)

The structural factor in Eq.~(\ref{26}) is equal to
\begin{eqnarray}
\gamma_{q}(0)= \alpha^{-1} {\eta\over 1+\eta^2 }
\left( 1 - {\cos(\eta q a)\over  \cosh(qa)}\right)
\label{35}
\end{eqnarray}

The dependencies  $\Delta v/v$ on $q$ are
shown in Fig.2 at
$\nu_f=2/5,3/7,4/9$ ($N=2,3,4$) for the lowest odd elastic mode.
The parameters  $c_b/c_l=1.5$, $2 a=1500 \AA$
are used (other parameters are the same ones).
We see, that the function of $\Delta v(q)$ oscillates and the
period of oscillation becomes smaller when $\nu_f$ approaches
to $\nu =1/2$ (it reflects the fact that the effective magnetic
length is increased).

\begin{figure}
\narrowtext
\centerline{\epsfig{figure=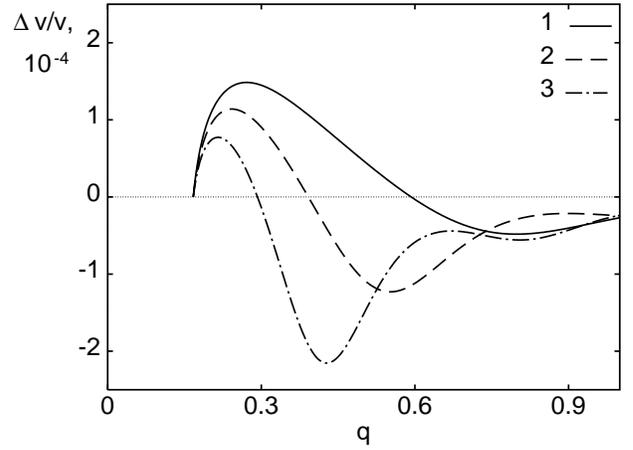,width=8cm}}
\vspace{0.5cm}
\caption{Velocity shifts of the lowest  odd mode vs the wave vector
for the monolayer electron
system at $c_b/c_l=1.5$.
1 - $\nu =2/5$;
2 - $\nu =3/7$;
3 - $\nu =4/9$;
$q$ is in $\sqrt{4 \pi n_0}$ units.  }
\label{fig2}
\end{figure}

\begin{figure}
\narrowtext
\centerline{\epsfig{figure=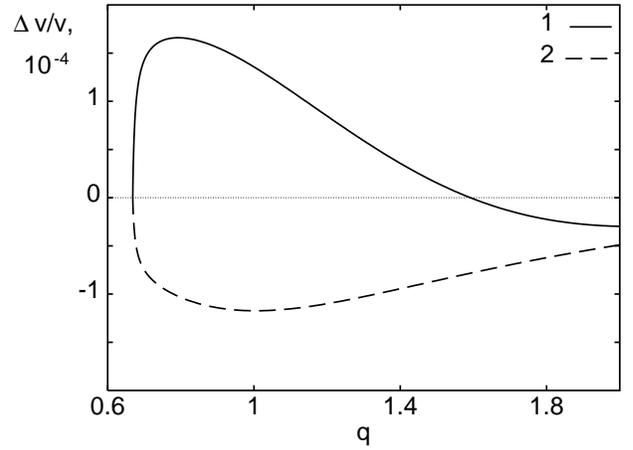,width=8cm}}
\vspace{0.5cm}
\caption{Velocity shifts of the third even mode vs the wave vector
for the double layer electron
system at $c_b/c_l=1.5$. 1 - $\varphi =4$, $s=0$; 2 - $\varphi=2$, $s=2$.
$q$ is in $\sqrt{4 \pi n_0}$ units.  }
\label{fig3}
\end{figure}

The absolute value of the velocity shift is much higher that
for the previous case. The effect is lowered at larger $a$.

If the heterostructure contains two electron layers at
$z=\pm d_0/2$,
then the phase transition between
$\psi=4$, $s=0$ and $\psi=2$, $s=2$ states is accompanied
by the shift of the velocity of the even mode only, as for the
previous case. The maximal value of the effect takes place for one
of higher harmonics (we specify the case of $d_0\ll 2 a$).

The structural factor is equal to
\begin{eqnarray}
\gamma_{q}(d_0/2)= -\gamma_{q}(-d_0/2)=\alpha^{-1} {\eta\over 1+\eta^2 } \cr
\times
\left({\sin(\eta q a)\sinh(q d_0/2)\over  \cosh(qa)}-
\sin(\eta q d_0/2)\right)
\label{36}
\end{eqnarray}
The dependencies  of $\Delta v/v$ vs $q$
are shown in Fig.3 for the third even harmonics
($d_0=300 \AA$, $2 a=1500 \AA$).

Thus, the measurements of the velocity shifts for the non-uniform plane
elastic modes of the waveguide type propagating in heterostructures
$AlGaAs-GaAs$ can be considered as a possible way of studying  the
dynamical properties of the monolayer and double layer quantum
Hall systems.
Unlike the case of the surface acoustic wave propagation,
the restriction on maximal values of the wave vectors,
for which the velocity shift can be observed
experimentally, does not take place for the waveguide modes.
In case of study of incompressible fractional Hall states,
the effect can be used for measuring the
dependence of the effective magnetic length
on the filling factor  and for observing the phase transitions in
the double-layer systems under the interlayer distance variation

\end{multicols}
\end{document}